# Limitations on Activation of High Dose Ge Implants in $\beta$-Ga$_2$O$_3$

Tianhai Luo[1], Katie R. Gann[1,4], Cameron A. Gorsak[1], Ming-Chiang Chang[1,5], Prescott E. Evans[2,3], Thaddeus J. Asel[2], Hari P. Nair[1], R. B. van Dover[1] and Michael O. Thompson[1]

[1]Department of Materials Science and Engineering, Cornell University, Ithaca, NY 14853, USA

[2]Air Force Research Laboratory, Wright-Patterson AFB OH 45433, USA

[3]Core4ce, Beavercreek OH 45324, USA

[4]U. S. Naval Research Laboratory, Washington, DC 20375, USA

[5]National Institute of Standards and Technology, Gaithersburg, Maryland 20899, USA

**Abstract**

Among ultrawide bandgap semiconductors, β-Ga$_2$O$_3$ is particularly promising for high power and frequency applications. For devices, n-type concentrations above >10$^{19}$ cm$^{-3}$ are required. Ge is a promising alternative n-type dopant with an ionic radius similar to Ga. Homoepitaxial (010) β-Ga$_2$O$_3$ films were implanted with Ge to form 50 and 100 nm box concentrations of 3×10$^{19}$ cm$^{-3}$ and 5×10$^{19}$ cm$^{-3}$, with damage ranging from 1.2 to 2.0 displacements per atom. For lower damage implants, optimized anneals in ultrahigh purity N$_2$ at 950-1000 °C for 5-10 minutes resulted in $R_S$ of 600-700 Ω/□, mobilities of 60-70 cm$^2$/Vs, and Ge activation of up to 40%. For higher damage implants, activation dropped to 23% with similar mobilities. Ge diffusion, measured by secondary ion mass spectrometry, showed formation of a Ge "clustering peak" with a concentration exceeding the initial implant following anneals in N$_2$ or O$_2$ at 950-1050 °C. Beyond this peak, minimal Ge diffusion occurred for N$_2$ anneals at 950 °C, but at 1050 °C non-Fickian diffusion extended to >200 nm. Electrical activation data suggests that clustered Ge is electrically inactive. To understand Ge clustering, several samples were characterized by synchrotron x-ray diffraction. Second-phase precipitates were observed in as-implanted samples which then fully dissolved after furnace annealing in N$_2$ at 1050 °C. Diffraction peaks suggest these implant-induced precipitates may be related to a high pressure $Pa\overline{3}$ phase of GeO$_2$, and may evolve during anneals to explain the Ge clustering. Ultimately, we believe Ge clustering limits activation of implanted Ge at high concentrations.

Ga$_2$O$_3$ has emerged as a promising material for next generation power electronics due to its ultra-wide bandgap.[1] Among the wide range of polymorphs, the thermodynamically stable β-phase can be melt-grown and is recognized as the most useful for power electronic devices. To realize its full potential, achieving low resistance metal contacts is critical, and requires high n-type doping at a level of 2-10×10$^{19}$ cm$^{-3}$.[2,3] Among potential n-type dopants, Si has been the most widely studied, though Ge and Sn also remain potential candidates. Germanium, in particular, has significant potential as a donor in Ga$_2$O$_3$ due to its similar ionic radius to Ga in tetrahedral sites (39 pm for Ge versus 47 pm for Ga, compared to Si at 26 pm),[4] and has a donor ionization energy comparable to Si.[5]

Ion implantation enables selective area doping with precise control over active dopant concentrations. Thermal annealing is subsequently required to recover the crystal structure and electrically activate dopants. While β-$Ga_2O_3$ has a high tolerance to radiation damage, resisting amorphization at room temperature,[6] high-dose implantation can introduce significant damage, including inducing a transformation to the γ-phase.[7–10] At high doses, only moderate activation of ion implanted Ge has been demonstrated.[11,12] Tetzner et al. implanted Ge into (100) β-$Ga_2O_3$, obtaining a 100 nm box profile of $5×10^{19}$ $cm^{-3}$. After pulsed thermal annealing to 1200 °C, a minimum sheet resistance of 590 Ω/□ was obtained with a mobility of 58 $cm^2$/Vs and an activation efficiency of 19.2% (relative to implant dose).[11] Spencer et al. similarly implanted Ge in (001) β-$Ga_2O_3$ for a 100 nm box profile of $3×10^{19}$ $cm^{-3}$. After annealing at 925 °C for 30 minutes, a sheet resistance of 926 Ω/□ was obtained with a mobility of 78 $cm^2$/Vs. [12]

Given the isoelectronic nature of Si and Ge, thermal annealing of Si and Ge implants is expected to be similar, though slightly higher temperatures might be expected for the larger Ge ions. Based on Si implant studies, control of defects within β-$Ga_2O_3$ during anneals is critical for high activation. Gann et al. showed that Si implants are sensitive to the annealing ambient as oxygen impacts defect formation, dopant activation, and diffusion. Partial pressures of oxidizing species must be reduced to $P_{O2}$ <$10^{-7}$ bar and $P_{H2O}$ <$10^{-8}$ bar for concentrations approaching $10^{20}$ $cm^{-3}$,[13] and suppression of gallium vacancy formation is believed to be the crucial factor for maximum electrical activity with minimal diffusion.[14,15]

In this work, we study the lattice damage, annealing, diffusion, and activation of Ge dopants following implantation to form 50 or 100 nm box profiles at concentrations from 3 to $5×10^{19}$ $cm^{-3}$.

Fe-doped (010) β-$Ga_2O_3$ substrates (23×25 mm) were obtained from Novel Crystal Technology. To avoid Fe diffusion from the substrate, a 500 nm unintentionally doped (UID) buffer layer was grown by metalorganic chemical vapor deposition (MOCVD) using conditions similar to previously reported.[15] Prior to implantation, a 20 nm $SiO_2$ cap was deposited by atomic layer deposition to ensure that the nominal implant doping box profile starts at the $Ga_2O_3$ surface. A single 23×25 mm substrate was diced into sections for three Ge implant conditions to ensure nominally identical samples. Using four energies (Table 1), 50 nm and 100 nm deep box-like concentration profiles were formed to target concentrations of 3 and $5×10^{19}$ $cm^{-3}$, with maximum implant damage ranging from 1.2 to 2.0 displacement per atom (DPA). After implant, each section was diced into final roughly 5×5 mm samples for annealing experiments. As needed, the $SiO_2$ cap was removed using a one-minute etch in 6:1 buffered oxide etchant (BOE).

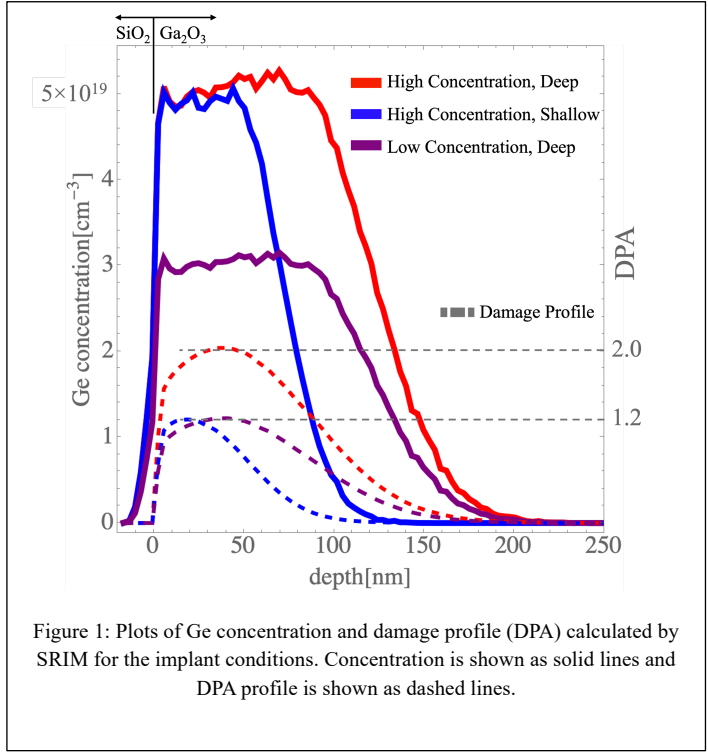

Figure 1: Plots of Ge concentration and damage profile (DPA) calculated by SRIM for the implant conditions. Concentration is shown as solid lines and DPA profile is shown as dashed lines.

Figure 1 shows the estimated Ge implant concentration (solid curves) as a function of depth determined by the Stopping Range of Ion in Matter (SRIM).[16] The dashed curves similarly show the estimated damage profiles (displacements per atom or DPA) for the three implant schedules.

These implant conditions (Table 1) were designed to cover a range of concentrations and depths commonly explored for Si implantation[13,17], and a range of calculated DPAs. In this manuscript, samples will be identified by the relative implant concentration and depth; H/D refers to high concentration ($5\times10^{19}$ cm$^{-3}$) and deep (100 nm), H/S to high concentration ($5\times10^{19}$ cm$^{-3}$) and shallow (50 nm), and L/D to low concentration ($3\times10^{19}$ cm$^{-3}$) and deep (100 nm). Samples H/D and H/S differ in the box thickness at similar Ge concentration. Sample L/D extends the implant to 100 nm with a total implant dose approximately the same as the H/S sample. The maximum damage for H/S and L/D samples are comparable near 1.2 DPA and will be collectively referred to as low-damage conditions, while the H/D sample has greater damage near 2.0 DPA and will be referred to as high-damage.

| Identifier | Box depth (nm) | Target box concentration (cm$^{-3}$) | Total dose (cm$^{-2}$) | Box dose (cm$^{-2}$) | Max DPA | Implant protocol Energy (upper entry, keV) Dose (lower entry, $\times 10^{13}$ cm$^{-2}$) | | | |
|---|---|---|---|---|---|---|---|---|---|
| High concentration, Deep (H/D) | 100 | $5.0 \times 10^{19}$ | $6.64 \times 10^{14}$ | $5 \times 10^{14}$ | 2.0 | 25 | 60 | 120 | 260 |
| | | | | | | 3.1 | 5.8 | 11 | 46.5 |
| High concentration, Shallow (H/S) | 50 | $5.0 \times 10^{19}$ | $3.87 \times 10^{14}$ | $2.5 \times 10^{14}$ | 1.2 | 25 | 60 | 70 | 160 |
| | | | | | | 3.1 | 5.6 | 1.0 | 29 |
| Low concentration, Deep (L/D) | 100 | $3.0 \times 10^{19}$ | $3.98 \times 10^{14}$ | $3 \times 10^{14}$ | 1.2 | 25 | 60 | 120 | 260 |
| | | | | | | 1.9 | 3.5 | 6.6 | 27.8 |

Table 1. Details for Ge implant conditions

Samples were annealed in an ultra-high vacuum compatible, quartz tube furnace under flowing 6N pure $N_2$ ($P_{O2} \leq 10^{-7}$ bar), further dried to $P_{H2O} \leq 10^{-8}$ bar, at temperatures from 950 to 1050°C for 5 or 10 minutes. As-implanted and annealed samples were further characterized by x-ray diffraction (XRD) to track lattice recovery and identify potential $\gamma$-phase precipitates often observed after implantation.[8,18–20] Both lab-based $2\theta$-$\omega$ diffraction, and glancing angle (2.5°) diffraction performed at the Cornell High Energy Synchrotron Source (CHESS), were utilized. Activated carrier concentrations and mobilities were determined by Hall measurement in the van der Pauw configuration using indium contacts soldered on the sample corners by hands. Dopant activation efficiency is defined as the ratio of measured sheet carrier density to the total implanted dose. Ge impurity profiles for as-implanted and annealed samples were measured by an IONTOF TOF-SIMS 5 instrument using a 2 kV $O_2^+$ sputter beam for positive secondary ion yield enhancement.

For Ge implant activation studies, samples were annealed under the optimized conditions determined for Si implants at comparable concentrations, namely $P_{O2}<10^{-7}$ bar, $P_{H2O}<10^{-8}$ bar, 950 °C and 10 minutes.[13] The $SiO_2$ cap, deposited prior to implants, was retained during most of the anneals to minimize interaction with the annealing ambient. Figure 2 shows the sheet resistance, mobility, and activation efficiency for the two low-damage samples (L/D and H/S) annealed at 950 °C or 1000 °C for 5 or 10 minutes. Sheet resistances of 600-700 Ω/□ were obtained with mobilities of 60-70 $cm^2$/Vs, and carrier densities up to ~$2\times10^{19}$ $cm^{-3}$ corresponding to activation efficiencies of 35% to 40%. One set of samples from each implant condition was also annealed at 950 °C after removing the $SiO_2$ cap; data from these anneals are shown as the first pair of data points in Figure 2. For high-damage samples (higher dose with 2.0 DPA), comparable sheet resistances and mobilities were obtained but with an activation efficiency that declined to 23%. These data indicate only minimal changes in electrical properties between samples annealed with and without a cap. At higher Si implant concentrations approaching $1\times10^{20}$ $cm^{-3}$, Gann *et al.* did observe significantly lower sheet resistances for samples annealed with an $SiO_2$ cap at 950 °C for 5 minutes.[15] From density function theory (DFT) calculations, the formation energy for gallium vacancies decreases with increasing Fermi level leading to increased sensitivity of anneals on $P_{H2O}$ and $P_{O2}$.[21] Use of a capping layer may help limit the impact of $P_{H2O}$ and $P_{O2}$ in a less well-controlled ambient (See SI Figure S1) by reducing gallium vacancy formation.[15] SIMS measurements also confirm minimal diffusion of Si from the $SiO_2$ cap into the film, limiting any unintentional donors (see SI Figure 5).

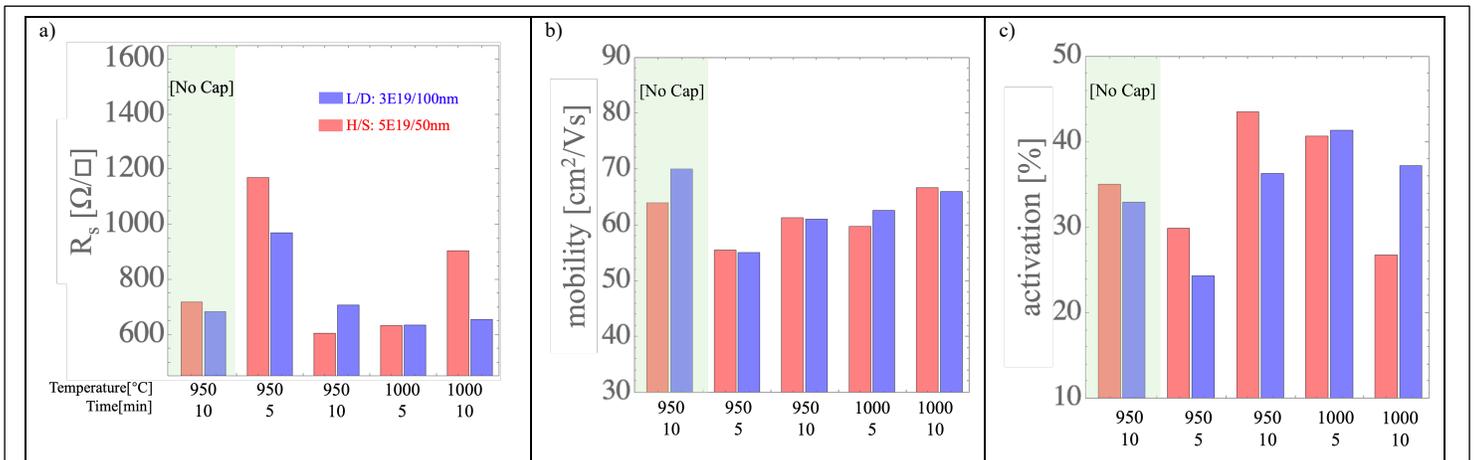

Figure 2. Plots of (a) $R_S$, (b) μ, and (c) activation rate, for L/D and H/S samples annealed sequentially without $SiO_2$ cap after an initial anneal with the cap. Joined points refer to sequential anneal.

SIMS measurements were used to track Ge diffusion with annealing time and temperature. Hall measurements only assess average charge transport through the entire film and will be influenced by dopant diffusion. Additionally, previous studies have shown that gallium vacancy defects accelerate substitutional cation diffusion,[22–26] especially during anneals in $O_2$.[23,24] Prior to SIMS, to establish the position of the $Ga_2O_3$ surface, any remaining $SiO_2$ cap was removed by a BOE etch followed by sputtering of a 50 nm $Al_2O_3$ layer. The β-$Ga_2O_3$ surface was then established at the point where the Ge concentration first increased. All SIMS spectra were scaled to the known implant dose.

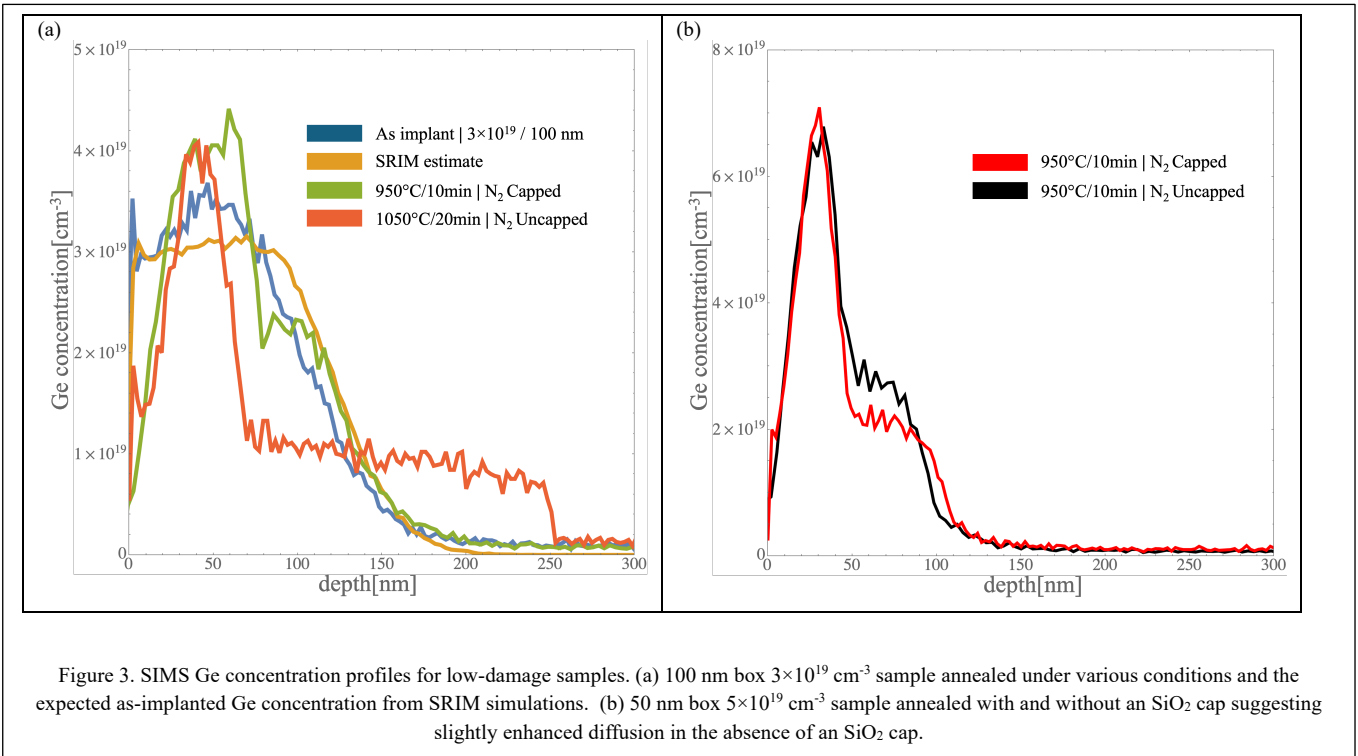

Figure 3. SIMS Ge concentration profiles for low-damage samples. (a) 100 nm box 3×10$^{19}$ cm$^{-3}$ sample annealed under various conditions and the expected as-implanted Ge concentration from SRIM simulations. (b) 50 nm box 5×10$^{19}$ cm$^{-3}$ sample annealed with and without an $SiO_2$ cap suggesting slightly enhanced diffusion in the absence of an $SiO_2$ cap.

Figure 3a shows SIMS concentration profiles of a 100 nm box low-damage sample as-implanted and after various anneals. The as-implanted profile agrees reasonably well with SRIM simulations. After annealing with an $SiO_2$ cap at 950°C for 10 minutes (green), Ge concentration at the surface decreases substantially with the formation of a Ge peak between 25 and 70 nm (referred to as a Ge clustering peak). The Ge in this peak was drawn from both the surface region and from 80-100 nm depths, leaving a plateau at 2.5×10$^{19}$ cm$^{-3}$ before following the decreasing concentration of as-implanted profile into the bulk. Beyond this plateau (>100 nm), Ge shows minimal diffusion with an estimated diffusivity <10$^{-16}$ cm$^2$/s ($D \approx L^2/4t$). Ge diffusion against the concentration gradient ("up-hill diffusion") into this clustering peak is unusual, resulting in a peak concentration above 4×10$^{19}$ cm$^{-3}$ compared to the initial box concentration of 3×10$^{19}$ cm$^{-3}$. After further annealing, to a total of 20 min at 1050 °C (red), the Ge clustering peak remains with an increased surface denuded zone (≈30 nm) and a peak concentration remaining above 4×10$^{19}$ cm$^{-3}$. Beyond the peak, however, significant non-Fickian diffusion forms a concentration plateau near 1×10$^{19}$ cm$^{-3}$ extending to 250 nm before decreasing sharply. This behavior is consistent with a strong concentration-dependent Ge diffusivity[23–25] estimated to exceed 2×10$^{-14}$ cm$^2$/s.

Similar behavior was observed for shallow low-damage implants. Figure 3b shows data for the 50 nm / $5\times10^{19}$ cm$^{-3}$ implant annealed at 950 °C for 10 minutes in $N_2$, with and without the $SiO_2$ cap. Ge again exhibits up-hill diffusion to a peak concentration near $7\times10^{19}$ cm$^{-3}$, well above the as-implanted concentration of $5\times10^{19}$ cm$^{-3}$. For this shallow implant, the Ge clustering peak occurs closer to the surface (30 nm) than for the deeper implant. Beyond the peak, a comparable plateau behavior was observed extending to over 100 nm at concentrations of $2\text{-}3\times10^{19}$ cm$^{-3}$. Diffusion during the uncapped anneal appears slightly increased ($\approx$10 nm), potentially linked with the expected higher gallium vacancy concentrations forming at the uncapped surface.

A second shallow low damage sample was also annealed, without an $SiO_2$ cap, under an $O_2$ ambient to determine if the Ge clustering was related to anneal ambient conditions. SIMS data (SI Figure S2a) show a similar Ge clustering peak at nearly the same location as for $N_2$ anneals. As expected for the high gallium vacancy concentrations associated with an $O_2$ ambient, diffusion beyond the peak was substantial with the plateau extending to >600 nm at low concentration. This sample was also non-conductive after annealing, consistent with defects formed during annealing under $O_2$.[27]

Based on the behavior under $O_2$ and $N_2$ annealing, we hypothesize that Ge in the cluster peak is inactive and only Ge in the plateau and tail regions contribute active carriers. SIMS Ge profiles were integrated from the start of the plateau (beyond the clustering peak) to 250 nm and compared with the electrically active dose determined from Hall measurements; results are summarized in Table 2. The SIMS determined Ge fractions past the cluster peak, 41.6-42.4% and 35.0-36.5% for the shallow and deep low-damage samples respectively, agree well with the measured activated rates of 41% and 35%. Some fraction of the Ge within the cluster is likely activated as well, but it is clear that increasing Ge activation beyond $\approx$40% can only be achieved if the Ge clustering is suppressed. Alternative models for the activated portions of Ge (*e.g.* the cluster peak itself or all Ge below the plateau level) were evaluated as well (SI Figure S3), but do not quantitatively match the observed activation efficiencies.

| Sample ID | Onset of Plateau (nm) | DPA @ Plateau Onset | Plateau Integral / Total Dose (%) | Measured Activation (%) |
|---|---|---|---|---|
| H/S, 950 °C/10min, #1 | 46 | $\approx$ 0.86 | 42.4 ($\pm$3) | 41 |
| H/S, 950 °C/10min, #2 | 46 | $\approx$ 0.86 | 41.6 ($\pm$4) | 41 |
| L/D, 950 °C/10min, #1 | 75 | $\approx$ 0.92 | 36.5 ($\pm$3) | 35 |
| L/D, 950 °C/10min, #2 | 75 | $\approx$ 0.92 | 35.0 ($\pm$3) | 35 |

Table 2. Quantitative analysis of Ge SIMS data for Ge dose beyond the clustering peak (plateau region to 250 nm). To establish reproducibility, measurements were repeated in two SIMS craters for each sample. The plateau onset, identified manually, has an estimated error of $\pm$1 data point $\pm$3 nm resulting in uncertainty in the Ge integral (as listed). Measured activation efficiency matches closely to this integrated Ge dose.

Identifying the physical mechanism leading to Ge clustering remains challenging. The peak position is nearly independent of annealing ambient ($O_2$ or $N_2$) and is located near the damage peak (Figure. 1). Such "up-hill diffusion" of Ge is similar to the redistribution of high dose implanted oxygen in Si during formation of buried $SiO_2$ layers for SIMOX (Separation by IMplant of OXygen),[28] but the dose of Ge here is far lower than that for $SiO_2$ formation ($10^{14}$ cm$^{-2}$ versus $10^{18}$ cm$^{-2}$). Other potential explanations include formation of second phase Ge-based precipitates (*e.g.* Ge or $GeO_2$ nanoparticles), or some form of an order-like enhancement in the local Ge occupancy on β-Ga$_2$O$_3$ Wykoff sites. Ge diffusion and precipitate formation in β-Ga$_2$O$_3$ have been previously reported. For example, during furnace anneals in an oxygen ambient, the diffusivity of Ge was observed to be nearly two orders of magnitude higher than either Si or Sn.[24] Similarly, during MBE growth of heavily doped films, surface segregation of Ge was observed even at growth temperatures as low as 550 °C.[29] And during implants at substantially higher doses and energy (2×10$^{16}$ cm$^{-2}$ at 650 keV), Fernández et al. reported formation of cubic Ge nanocrystals (Fd$\bar{3}$m) after annealing at 600 °C in air, with these nanocrystals dissolving at higher temperatures leaving nano-voids.[30] These data suggest a high diffusivity of Ge, and more critically a large positive enthalpy of mixing for Ge in β-Ga$_2$O$_3$ which would provide the driving force for the segregation.

To identify potential (epitaxial or non-epitaxial) precipitates, 2θ-ω XRD scans over extended angles, and synchrotron-based glancing angle XRD scans, were obtained. For as-implanted samples, 2θ-ω XRD scans (Figure S6) showed only formation of the commonly observed broad peak attributed to coherent γ-phase regions.[8,10,18,31] After annealing at 950°C for 10 min, scans showed full lattice recovery (SI Figure S6) with no features beyond β-Ga$_2$O$_3$ strain.[32,33] To identify any non-epitaxial second phase precipitates, synchrotron-based glancing-angle XRD scans were obtained for an as-implanted and a 1050 °C annealed high-damage H/D sample. Compared to conventional lab sources, the high x-ray flux from the synchrotron and the glancing angle geometry dramatically increased the sensitivity to polycrystalline precipitates in the film. Figure 4 shows 2D $q$-$\chi$ XRD patterns (transformed from acquired 2D x-ray images)[34] from as-implanted (a) and annealed (b) samples. In these images, diffraction from randomly oriented polycrystalline grains would be observed as uniform intensity lines at a fixed $q$ value across all $\chi$ angles.

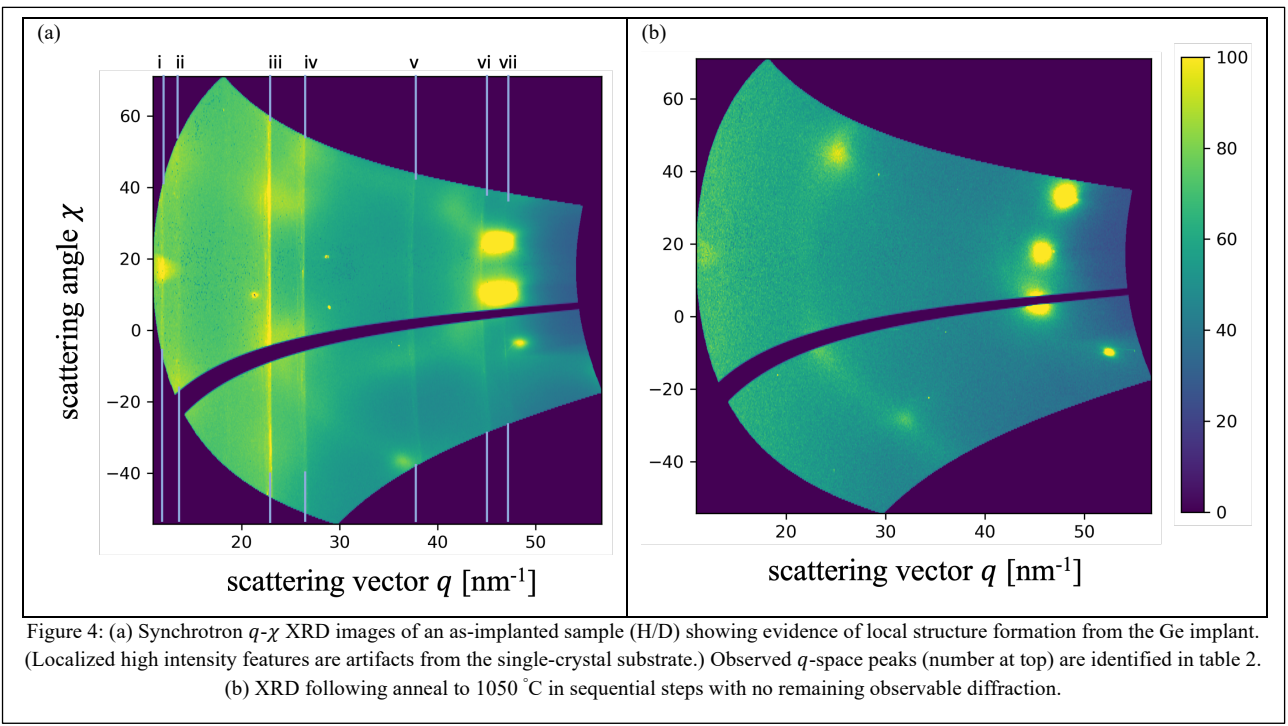

Figure 4: (a) Synchrotron $q$-$\chi$ XRD images of an as-implanted sample (H/D) showing evidence of local structure formation from the Ge implant. (Localized high intensity features are artifacts from the single-crystal substrate.) Observed $q$-space peaks (number at top) are identified in table 2. (b) XRD following anneal to 1050 °C in sequential steps with no remaining observable diffraction.

The as-implanted sample (Figure 4a) shows clear evidence of scattering from additional phase(s), as discussed below. However, after annealing (Figure 4b), any phase(s) present from the implant dissolved with no remaining evidence of non-epitaxial precipitates, similar to the dissolution of Ge nanoparticles observed by Fernández et al..[30] These data indicate that Ge clustering observed in the SIMS data is likely not related to second-phase precipitation, which is also consistent with lab-based $2\theta$-$\omega$ XRD scans showing full epitaxial recovery.[32,33]

In the as-implanted films, at least seven diffraction lines were identified (i through vii) with some showing limited evidence of splitting. These lines are listed in Table 3 with the scattering vector ($q$), qualitative intensity estimates (very weak to strong), and the corresponding diffraction plane spacing ($d$). These peaks reflect near-cubic symmetry but do not correspond to any known $Ga_2O_3$ polymorph, including the cubic $\gamma$-phase, nor to any reported Ge phases. The high-$q$ diffraction lines (ii-vii) are consistent with a simple cubic phase having a lattice constant of 4.73 ± 0.05 Å, as detailed in columns 5 and 6 of Table 3. DFT calculations predict a non-equilibrium (at ambient conditions) pyrite phase of $GeO_2$ ($Pa\bar{3}$ space group) forming at pressures above 70 GPa with a lattice constant of 4.4 Å.[35,36] The Materials Project includes this phase (mp-2633)[37] with a zero pressure lattice constant of 4.7 Å, and experimentally the phase has been observed at 126 GPa with a lattice constant of 4.33 Å.[38] With the exception of the first two lines, the diffraction data are consistent with precipitation of this pyrite $GeO_2$ phase during ion implantation; expected line intensities of $GeO_2$ diffraction are included in the last column of Table 3 and qualitatively match observed intensities. The two low-$q$ reflections (i and ii) are not immediately consistent, but may result from $\lambda/2$ harmonic contamination in the synchrotron beam, symmetry breaking allowing the forbidden reflections, and/or longer-range ordering of the lattice.

| Peak | Measured $q$ [nm$^{-1}$] | Intensity | d-spacing [Å] | Potential cubic reflection | Implied Lattice Constant (Å) | $GeO_2$ ($Pa\bar{3}$) relative reflection intensity |
|---|---|---|---|---|---|---|
| i | 11.8 | Weak | 5.324 | N/A | N/A | N/A |
| ii | 13.5 | Very Weak | 4.654 | 100 | 4.654 | forbidden |
| iii | 22.9 | Strong | 2.744 | 111 | 4.753 | 100 |
| iv | 26.5 | Medium | 2.371 | 200 | 4.742 | 53.1 |
| v | 37.8 | Medium | 1.662 | 220 | 4.701 | 29.8 |
| vi | 45.0 | Weak | 1.396 | 113 | 4.630 | 18.78 |
| vii | 47.1 | Very Weak | 1.334 | 222 | 4.621 | 12.5 |

Table 3. Details of the reflections from synchrotron XRD on H/D as-implanted sample. N/A is an abbreviation for "Not Applicable".

In contrast to the known commensurate $\gamma$-phase formation in the β-phase lattice, these precipitates are non-epitaxial but instead form randomly oriented grains. Formation of this (high-symmetry) phase during the ion-implant collision cascades would require both a high driving force (positive enthalpy of Ge mixing) and high Ge diffusivities, as suggested by earlier studies.[24,29,30] At the doses explored, $GeO_2$ precipitates could not exceed 0.1% volume fraction. Given this limit, and the limited Debye-Sherrer broadening suggesting relatively large precipitates, and the similar x-ray scattering of Ge and Ga, we suggest formation of $(Ge_{1-x}, Ga_x)O_{2-\delta}$ structures, potentially explaining the additional diffraction lines observed. Additional studies are required, but we believe that the observed Ge segregation after annealing is related to this initial formation of Ge-rich precipitates during the implant process.

In conclusion, Ge dopant activation and diffusion were studied for (010) epitaxial β-$Ga_2O_3$ films implanted to form 50 or 100 nm box profiles with concentrations of 3 or $5\times10^{19}$ cm$^{-3}$, and with maximum implant damage levels of 1.2 and 2.0 DPA. For low-damage samples (1.2 DPA), annealing under high purity $N_2$ ($P_{O2} \leq 10^{-7}$ bar, $P_{H2O} \leq 10^{-8}$ bar) at temperatures between 950 and 1000 °C activated Ge with sheet resistances of 600-700 Ω/□, mobilities of 60-70 cm$^2$/Vs, and activation efficiencies reaching over 40%. For high-damage samples (higher dose with 2.0 DPA), comparable sheet resistances and mobilities were obtained but with an activation efficiency that declined to 23%. Ge diffusion during annealing was studied by SIMS. For low-damage samples annealed in $N_2$ between 950 and 1050 °C, SIMS showed formation of a Ge clustering peak with peak Ge concentration in this peak of $4\times10^{19}$ and $7\times10^{19}$ cm$^{-3}$, exceeding the initial box implant concentrations of $3\times10^{19}$ and $5\times10^{19}$ cm$^{-3}$ respectively. Similar clustering was also observed after annealing in $O_2$. Beyond the peak, minimal Ge diffusion was observed for 10 minute anneals at 950°C. However, annealing at 1050°C for 20 minutes resulted in Ge diffusion to >250 nm, with the Ge profile forming a nearly constant concentration profile before dropping sharply. The integral of Ge concentration outside the cluster peak is shown to match the measured active carrier concentration, suggesting that Ge within the cluster peak is inactive. Synchrotron XRD of high-dose as-implanted samples showed formation of randomly oriented near-cubic precipitates, which we suggest may be $(Ge_{1-x}, Ga_x)O_{2-\delta}$ pyrite structures formed within the collision cascade as a result of high Ge diffusivity and a large positive enthalpy of mixing. After annealing to 1050 °C in $N_2$, these precipitates dissolve to concentration below the detection limit. Ultimately, Ge clustering is a fundamental limit to activation of Ge in β-$Ga_2O_3$, especially for high dose implants.

This research was supported by the Air Force Research Laboratory-Cornell Center for Epitaxial Solutions (ACCESS) under Grant No. FA9550-18-1-0529. The authors also acknowledge use of the Materials Solutions Network at CHESS (MSN-C) supported by the Air Force Research Laboratory under Award No. FA8650-19-2-5220, and the Cornell Nanoscale Facility, a member of the National Nanotechnology Coordinated Infrastructure (NNCI), which is supported by the National Science Foundation (Grant No. NNCI-2025233). C.A.G. acknowledges support from the National Defense Science and Engineering Graduate (NDSEG) fellowship.

The authors have no conflicts to disclose